\documentclass[review,sort&compress]{elsarticle}
\journal{Nucl. Instr. and Meth. in Phys. Res. A}

\usepackage{graphicx} \graphicspath{{figures/}}
\usepackage{amssymb}
\usepackage{lineno}

\def\kaos{{\sc Kaos}\@}

\begin{document}
\linenumbers

\begin{frontmatter}

  \title{Particle tracking in kaon electroproduction with
    cathode-charge sampling in multi-wire proportional chambers}

  \author[kph]{P. Achenbach\corref{cor}} \ead{patrick@kph.uni-mainz.de}
	\author[kph]{C.~{Ayerbe Gayoso}}
	\author[kph]{J.~C.~Bernauer}
	\author[kph]{R.~B\"ohm}
	\author[zagreb]{D.~Bosnar}
	\author[kph]{M.~B\"osz}
	\author[ljubl]{L.~Debenjak}
	\author[kph]{M.~O.~Distler}
	\author[kph]{A.~Esser}
	\author[zagreb]{I.~Fri\v s\v ci\'c}
	\author[kph]{M.~{G\'omez Rodr\'iguez de la Paz}}
	\author[zagreb]{M.~Makek}
	\author[kph]{H.~Merkel}
	\author[kph]{U.~M\"uller}
	\author[kph]{L.~Nungesser}
	\author[kph]{J.~Pochodzalla}
	\author[ljubl]{M.~Potokar}
	\author[kph]{S.~{S\'anchez Majos}}
	\author[kph]{B. S.~Schlimme}
	\author[ljubl]{S.~\v Sirca}
	\author[kph]{M.~Weinriefer}

  \cortext[cor]{Corresponding author. Tel.: +49-6131-3925831; fax:
    +49-6131-3922964.}

  \address[kph]{Institut f\"ur Kernphysik, Johannes
    Gutenberg-Universit\"at Mainz, Germany}
  \address[zagreb]{Department of Physics, University of Zagreb,
    Croatia} 
  \address[ljubl]{University of Ljubljana and Jo\v zef
    Stefan Institute, Ljubljana, Slovenia}

  \begin{abstract}
    Wire chambers are routinely operated as tracking detectors in
    magnetic spectrometers at high-intensity continuous electron
    beams. Especially in experiments studying reactions with small
    cross-sections the reaction yield is limited by the background
    rate in the chambers.  One way to determine the track of a charged
    particle through a multi-wire proportional chamber (MWPC) is the
    measurement of the charge distribution induced on its cathodes. In
    practical applications of this read-out method, the algorithm to
    relate the measured charge distribution to the avalanche position
    is an important factor for the achievable position resolution and
    for the track reconstruction efficiency.  An algorithm was
    developed for operating two large-sized MWPCs in a strong
    background environment with multiple-particle tracks.  Resulting
    efficiencies were determined as a function of the electron beam
    current and on the signal amplitudes. Because of the different
    energy-losses of pions, kaons, and protons in the momentum range
    of the spectrometer the efficiencies depend also on the particle
    species.
  \end{abstract}


\end{frontmatter}

\section{Introduction}

Since long, high-resolution magnetic spectrometers for charged
particles have been used in nuclear and particle physics. Combined
with suitable coordinate detectors for tracking, such instruments can
achieve the most accurate measurement of kinematic
quantities~\cite{Blomqvist1989}. For focusing spectrometers the
measurements can be done in an detection area out of the line-of-sight
to the target that is shielded to some extend from background
particles.  At the two currently operating high-energy, high-intensity
continuous electron beam accelerators MAMI in Mainz, Germany, and
CEBAF at Jefferson Lab (JLab), Virginia, wire chambers are routinely
operated as tracking detectors in magnetic
spectrometers~\cite{Baker1995,Blomqvist1998,Fissum2001}.

During the last decade, a class of strangeness production experiments
represented by the $(e,e'K^+)$ reaction were realized at the
spectrometer facilities of both laboratories. The small cross-sections
encountered in these reactions require high luminosities and in case
of hypernuclear reactions the use of nuclear targets up to the
medium-mass region ($A\leq$ 52). To maximise hypernuclear yields, the
spectrometers are operated at minimum forward angles, where background
rates increase by several orders of magnitude, severely interfering
with the operation of the wire chambers.

In the pilot $(e,e'K^+)$ hypernuclear experiment at JLab, E89-009, a
very high rate of electrons associated with bremsstrahlung dominated
the background in the electron spectrometer and a very high rate of
positrons from Dalitz pairs dominated the particle flux in the kaon
spectrometer, so that beam currents were limited to below 1\,$\mu$A,
giving an experimental luminosity, $\cal L$, of approximately 0.001
$\times$ 10$^{36}$\,cm$^2$s$^{-1}$~\cite{Miyoshi2003}.  Even after a
decade of optimizations the wire chambers of the third generation
experiment at JLab, E05-115 (HKS-HES), experienced multiple tracks for
a medium-mass target of 150\,mg$/$cm$^2$ and beam currents of several
$\mu$A ($\cal L$= 0.09 $\times$ 10$^{36}$\,cm$^2$s$^{-1}$) that could
not be resolved by the standard tracking analysis with a track
reconstruction efficiency higher than $\sim$
50\,\%~\cite{Gogami2010:SPHERE}.

The operation of multi-wire proportional chambers (MWPCs) 
in the \kaos\ spectrometer at MAMI is discussed in Section~\ref{sec:MWPC}. 
It is described how the centre-of-charge analysis
method, well-established in single-track events, had to be adapted
to the multiple-track events. The strong
variation in the charge distributions and the ambiguity introduced by
a large number of charge clusters per event has lead to the
development of a cluster and track-finding algorithm for the MWPCs,
which is presented in Sections~\ref{sec:clusters} and
\ref{sec:tracks}. This algorithm succeeded in resolving the
multiple-track ambiguities with high efficiency. It was in-beam tested
with a set of dedicated efficiency counters which allowed to check the
validity of possible tracks. Results of the efficiency measurements
are presented in Section~\ref{sec:efficiencies} and the application of
the method is discussed in Section~\ref{sec:discussion}.

\section{Operation of the MWPCs}
\label{sec:MWPC}

Crucial requirements for the wire chamber operation in kaon
electroproduction measurements are (i) high efficiency for minimum
ionizing particles and (ii) tracking capability for luminosities above
5 $\times$ 10$^{36}$\,cm$^2$s$^{-1}$.  The limit for the second
requirements depend strongly on the wire chamber geometry. It is known
that slow-moving ions from the avalanches around the anodes form space
charges that reduce the field.  The effective voltage drop for a given
particle rate increases with the anode pitch, with the distance
between anode and cathode plane, and with the gas
gain~\cite{Mathieson1991}. Modern developments aim for MWPC that
resolve track coordinates with a spatial resolution of about
100\,$\mu$m in a very high-rate background up to a few MHz$/$cm$^2$.
Such chambers have small chamber dimensions and are operated with low
gas gain.

At the Institut f\"ur Kernphysik in Mainz, Germany, two large-sized
MWPCs are operated by the A1 Collaboration as tracking detectors in
the \kaos\ spectrometer's hadron arm~\cite{Achenbach2010:HypX}. They
were used before in several beam-times at GSI for trajectory
reconstruction at relativistic heavy ion experiments, detecting
successfully trajectories of pions and kaons as well as protons and
heavy nuclear fragments~\cite{Stelzer1991,Senger1993}. Contrary to
conventional MWPCs the multiplication process in these chambers is
divided into two steps, following a concept first described by Breskin
and co-workers~\cite{Breskin1979}. During the electron scattering
experiments at MAMI the distribution of charge states of the ionizing
particles as well as particle fluxes differ greatly from the situation
at GSI. Typically the detector packages of spectrometers at electron
machines are heavily shielded.  At MAMI, the the shielding houses of
the three-spectrometer facility consist of 40\,cm thick boron carbide
loaded concrete walls covered with a 5\,cm thick lead layer on the
inside. They weigh about 110 tons each~\cite{Blomqvist1998}.  For the
shielding of the \kaos\ spectrometer polyethylene shields with 15\%
diboron trioxide by weight were used in combined system of shielding
walls and ceiling covering 20\,m$^2$. The walls comprise a 10\,cm
thick neutron shield and a 5\,cm thick lead layer.  At the
\kaos\ spectrometer the chambers, placed behind the analysing dipole
magnet, are shielded by polyethylene walls and ceiling with 15\%
diboron trioxide by weight covering 20\,m$^2$. The walls comprise a
10\,cm thick neutron shield and a 5\,cm thick lead layer.  However,
when a continuous-wave electron beam of several $\mu$A current is
delivered to solid-state or cryogenic liquid targets situated in front
of the dipole the chambers experience multiple tracks in almost every
event due to high electromagnetic background radiation levels.

\begin{figure}
  \centering
  \includegraphics[width=\columnwidth]{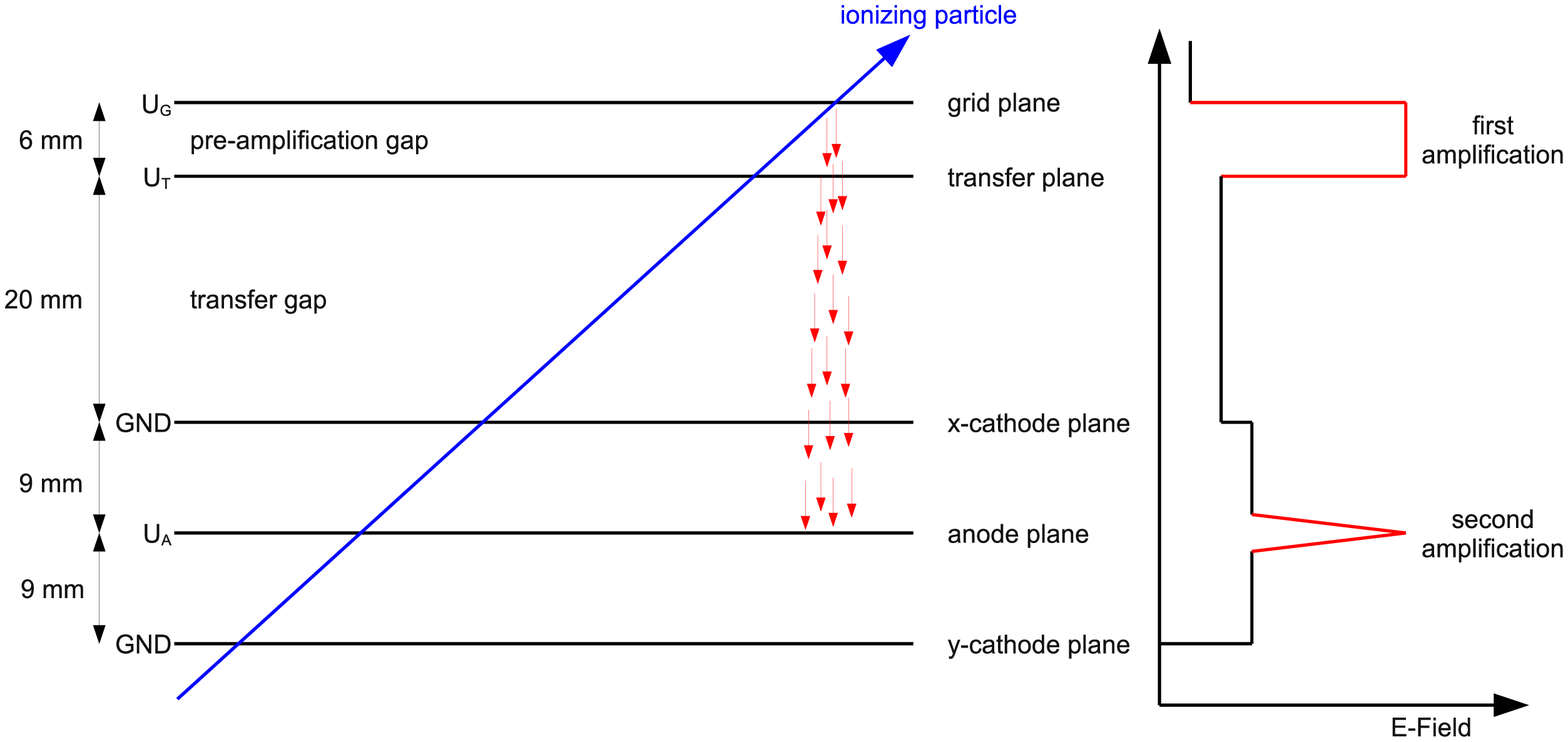}
  \caption{Schematic layout of the MWPCs with two-stage gas
    amplification. Each chamber consists of a plane of anodes (\o\ =
    20\,$\mu$m gold-plated tungsten wires, 2\,mm spacing),
    symmetrically sandwiched between two orthogonal planes of cathodes
    (\o\ = 50\,$\mu$m gold-plated tungsten wires, 1\,mm spacing) along
    with two conducting grids, providing two gaps in front of the wire
    planes. On the right side a typical distribution of the electric
    field is shown. Electrons liberated in the pre-amplification gap
    are amplified before drifting through the transfer-gap, the
    transferred charge is amplified a second time at the anodes.}
  \label{fig:MWPC-schematic}
\end{figure}

The two MWPCs, labelled M and L, have an active area of 1\,190
$\times$ 340\,mm$^2$ each. The chambers are schematically depicted in
Fig.~\ref{fig:MWPC-schematic}. They consist of a plane of anodes
(\o\ = 20\,$\mu$m gold-plated tungsten wires, 2\,mm spacing),
symmetrically sandwiched between two orthogonal planes of cathodes in
$x$- and $y$- direction (\o\ = 50\,$\mu$m gold-plated tungsten wires,
1\,mm spacing) along with two meshes of woven fabrics of plastic
coated with a nickel layer, making up two planar electrode structures,
the grid ($G$) and the transfer ($T$) plane and two gaps, the
pre-amplification gap and the transfer-gap. The wires of the anode
plane are running in diagonal direction, making an angle of 45$^\circ$
with either cathode direction. Typical potentials applied to these
electrodes are: $U_G =$ $-$9.1\,kV, $U_T =$ $-$2.0\,kV, $U_A =$
$+$4.0\,kV, with the cathodes grounded.

Despite modern alternatives like micro-pattern gas detectors, the use
of multi-wire proportional chambers (MWPCs) is still an economical way
to cover a large area with tracking devices for charged particles.
MWPCs are especially attractive when the particle track can be
recorded from a single chamber in two orthogonal dimensions and when
the required resolution is not significantly finer than FWHM
$\lesssim$ 1\,mm.

To determine the particle track the measured charge distributions
induced on the cathode wire planes of the MWPCs are analysed.  Of the
several known methods for the bi-dimensional read-out of MWPCs, the
induced-charge sampling is well established since the
1970s~\cite{CharpakSauli1973}. However, even in single-track
operation, the choice of analysis method for the induced charge
distributions on the cathode strips or wires has consequences on the
position resolution and the track reconstruction
efficiency~\cite{Awaji1982}. The method can give good spatial
resolution, but is compromised for multiple particles in one event as
the signals can get partially integrated, the wide charge
distributions can overlap, and peaks can get distorted.

Five cathode wires are connected together to one channel and
are brought to one charge-sensitive preamplifier making a total of 240
analogue channels in $x$- and 70 analogue channels in $y$-direction.
The preamplifiers provide a bipolar output with a negative amplitude
of up to 2.5\,V. The pulse width is $\sim$ 2\,$\mu$s. The signals are
digitized by an ADC card addressed by a freely programmable transputer
module, mounted directly on the chamber. The ADC converts the signal
within 1.34\,$\mu$s into 8 bit. Sixteen of the ADC channels are read
out and processed by one transputer. The transputer network system is
connected to a multi-link card inside a front-end computer.  As the
FWHM of an induced cathode signal is known to be nearly equal to twice
the anode-to-cathode gap, this is 2 $\times$ 9\,mm, clusters of
signals in 3--5 channels are observed.

Particles from the target cross the MWPC planes with an angle of about
(50 $\pm$ 20)\,$^\circ$ to the normal. The peak position of charges
transferred from the pre-amplification gap to the anodes is assumed to
coincide with the impact point of the trajectory of the ionizing
particle with the grid plane. This position does not depend on the
angle of incidence of the particle's trajectory, whereas the shape of
the charge distribution is influenced.

The chambers are filled with a gas mixture of \{84\,\% Ar, 9\,\%
CO$_2$, 7\,\% C$_4$H$_{10}$\} in volume concentration. Calibrated
flow controllers are used to control the amounts of the three gas
components.

Details of the induced charge distributions depend on experimental
parameters like track angles, particle species and velocities, and
multiplicities, but also on working conditions like gas mixture, high
voltages, and the integration gate width and timing.  Therefore it is
of practical importance to apply an algorithm which gives the best
position resolution and highest track reconstruction
efficiency. However, the choice of the analysis method depends heavily
on the experimental requirements.

\section{Cluster Analysis}
\label{sec:clusters}

\begin{figure}
  \centering
  \includegraphics[width=\columnwidth]{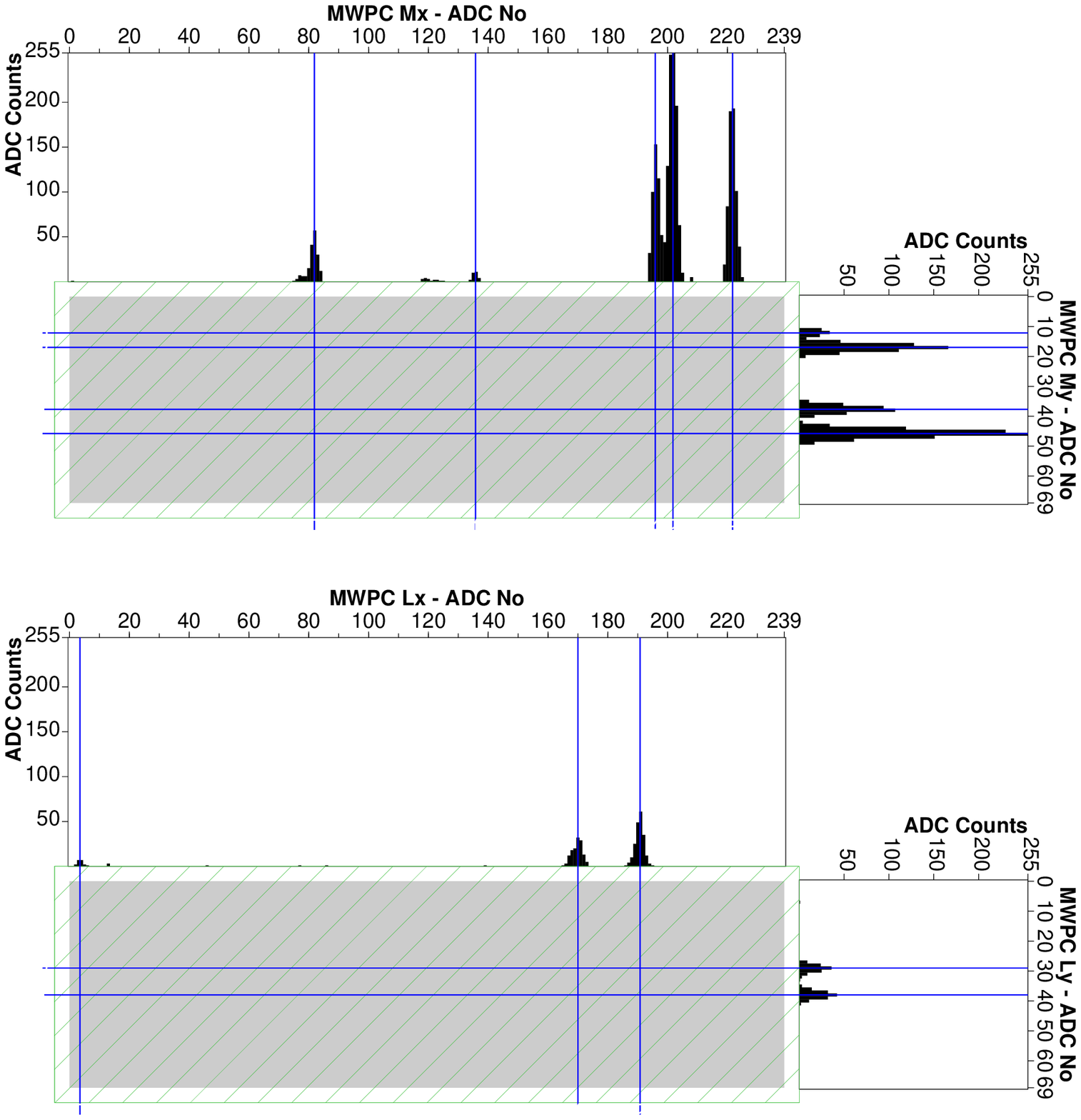}
  \caption{Charge distribution in both MWPCs of an event taken with
    2\,$\mu$A beam current, corresponding to luminosities of 2.5
    $\times$ 10$^{36}$\,cm$^2$s$^{-1}$. The anode wires are shown
    schematically as diagonal lines, cathode wires run horizontally
    and vertically. The grey shaded area corresponds to the active
    region of a chamber. Peak positions are determined by calculating
    the centre-of-charge from a restricted number of cathode channels
    and are indicated by vertical and horizontal lines in the display.
    The peak-finding algorithm assumes a typical peak FWHM of 3--5
    cathode strips and allows for asymmetric tails inherent to the
    chamber geometry.}
  \label{fig:MWPCevent164}
\end{figure}

\begin{figure}
  \centering
  \includegraphics[width=\columnwidth]{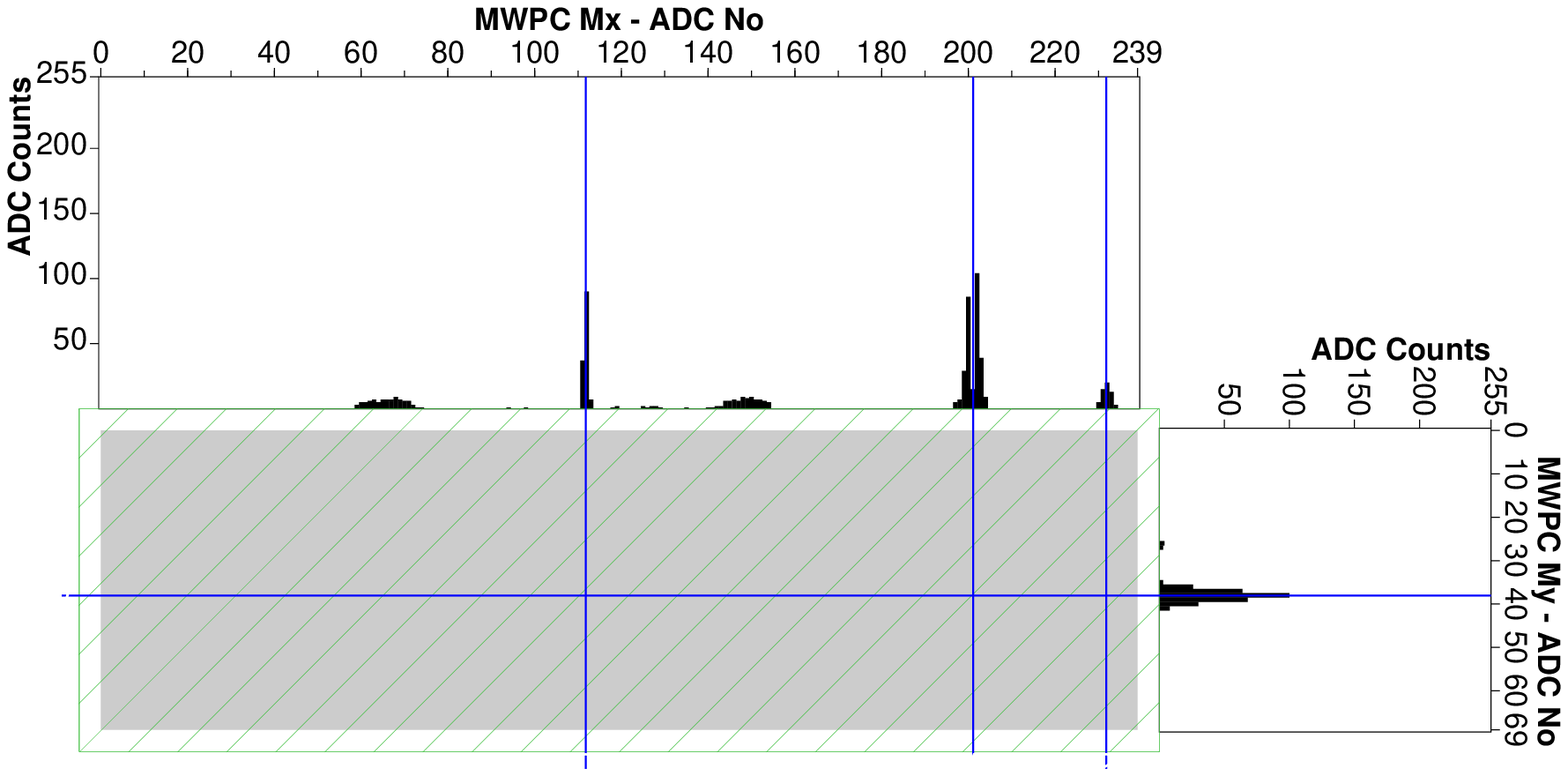}
  \caption{Charge distribution in MWPC~M showing satellite charges on
    both sides of a needle-like peak in the $x$-plane.  Such
    distributions are formed by multiple particles traversing the
    chamber at different times, so that the integration time window is
    offset and the integration is sensitive to charges induced on the
    cathode strips by charges flowing out of the anode wires. The
    coupling is seen by comparing the end-points of the diagonal anode
    wire crossing the vertical line of the needle peak to the position of
    the satellite peaks. The peak-finding algorithm rejects those
    satellites on the basis of their shape.}
  \label{fig:MWPCevent0}
\end{figure}

Fluctuations in the primary ionisation and electron diffusion along
the drift path lead to non-Gaussian charge distributions induced on
the cathodes.  The distributions are further deteriorated by noisy
channels, by multiple tracks in the chamber, and by induced signals in
the electronic chain. To relate the induced charge distributions to
the avalanche position, (i) the geometrical parameters of the chamber,
(ii) the signal-to-noise ratio, and (iii) physical processes of the
ionization and avalanche formation had to be taken into
account. Charge distributions in both MWPCs taken with 2\,$\mu$A beam
current on a 5\,cm liquid hydrogen target, corresponding to
luminosities of 2.500 $\times$ 10$^{36}$\,cm$^2$s$^{-1}$, are shown in
Figs.~\ref{fig:MWPCevent164} and \ref{fig:MWPCevent0}. Peak positions
as indicated by vertical and horizontal lines in the display were
determined by the cluster algorithm discussed in this Section.

The raw data are 8-bit ADC values, $Q_{i}$, for each read-out channel,
$x_i$, in the $x$- and $y$-plane.  As preamplifiers and ADCs vary
along the planes, each combination is assigned a gain factor $c_{i}$
and a pedestal value $p_{i}$. In few cases, where individual channels
showed misbehaviour, caused {\em e.g.}\ by wire ageing, their
amplitudes were interpolated from both neighbouring channels. Then
``clusters'' are defined by a group of neighbouring channels with
detected charges. A cluster has to consist of at least 2 channels and
could include a single channel below pedestal.  At first each cluster
is characterized by its detected charge, $Q = \sum_{n} Q_{i}\cdot
c_{i}$, its channel multiplicity, $n$, its centre-of-charge, $\bar{x}
= (\sum_{n} Q_{i} \cdot c_{i} \cdot x_{i})/(\sum_{n}Q_{i} \cdot
c_{i})$, its width, $\sigma = \sqrt{\sum_{n} \left| X_{i}-\bar{x}
  \right|^{2} Q_i/Q}$, and its largest amplitude, $Q_{max}$, at
channel $x_{max}$.

Because of the two amplification regions, as shown in
Fig.~\ref{fig:MWPC-schematic}\,(right), and the large crossing angles
of the particles, the liberated charges that are drifting from the
pre-amplification plane will be displaced in the anode plane with
respect to the weaker direct signal. The two separated charge
localisations differ by a distance $x = d \tan\theta$, $d$ being
35\,mm. For a typical angle of $\theta = 55^{\circ}$ this distance was
$x \cong 50$\,mm corresponding to 10 channels.  As the amplification
of the direct signal was only partial, and the signal was induced on
the cathodes at a much earlier time and thus got integrated only
partially, a second peak in the charge distribution was not often
observed, however, an asymmetry in the shape of the peak was regularly
present in the signals. Because the drift time was regulated by the
gas mixture and the high voltages, and the integration gate could get
matched to this drift time, it was possible to minimize the appearance
of double peaks and to reduce variation in cluster shapes.  In
Fig.~\ref{fig:MWPCevent164} the peaks in the $x$-plane of MWPC~M show
features related to the two-stage amplification.

During the experiments sometimes charge distributions were observed,
where satellite peaks were accompanying the physical peak, that was
produced by a heavily-ionising light particle.
Fig.~\ref{fig:MWPCevent0} shows a charge distribution in the $x$-plane
of MWPC~M with satellite charges on both sides of a needle-like
peak. The source of these peaks can be understood by the anode plane
construction mounted in between the cathode planes and was identified
as a capacitive coupling between anode and cathode at the stesalite
frame which have a dielectric constant $\epsilon_r \approx$ 6, see
discussion in~\cite{Baltes1992:GSI91}. A strong correlation exits
between the vertical position of the physical peak and the relative
position of the satellites in the $x$-plane.  Typical distances
between the satellites are 82--86 cathode channels reflecting the
height of the frame.  The satellite signals were not be present in
single-track events after a fine adjustment of the integration gate,
but could not be avoided in multi-track events, when the integration
gate was offset in respect to the timing of the second traversing
particle, and the integration became sensitive to charges induced on
the cathode strips by charges flowing out of the anode wires.


Each cluster is further analysis for its internal structure 
to identify possible peak positions as
follows: A truncated cluster is created around the channel with the
charge maximum, $x_{max}$. It extends to the left and to the right
for 2 channels, if the charge values for these channels
are below a minimum of 90\,\% of the charge in $x_{max}$. The
truncated cluster extends at most up to the limits of the original
cluster or up to a channel with only a minimum of charge.  For
clusters that reach the 8-bit maximum of the ADC the counting of
channels starts at the left and right limits of the saturation
plateau.  If no such structure is found the cluster is discarded. This
procedure ensured that a truncated cluster corresponds to a peak-like
structure.  Next, the summed charge in the truncated cluster is
compared to the charge outside the peak region. If the latter is
larger than 30\,\% in the $x$-planes of the MWPC or larger than 1\,\%
in the $y$-planes of the MWPC and the outside region is wider than the
required minimum for a cluster, a separate cluster is created. The
difference between the $x$- and $y$-planes is attributed to the
different appearances of pick-up charges on the cathode wires. The
extensions of the new cluster is depending on the distribution of
charges. For a large charge close to the original peak, {\em i.e.}\
within 2 channels, the new cluster extends to the peak region,
otherwise to the border region. This construction avoids the creation
of extra clusters when fluctuations appear in the tails of the
original peak.

\section{Track Reconstruction}
\label{sec:tracks}

In the analysis of the data all clusters are combined to form possible
particle tracks. With $n$ clusters $n^2$ possible track points and
$n^4$ possible tracks through both chambers are generated. This
ensemble is classified according to a set of track quality factors,
ranging from 0 (excluded) to 1 (highest quality).

\begin{figure}
  \centering	 
  \includegraphics[width=\columnwidth]{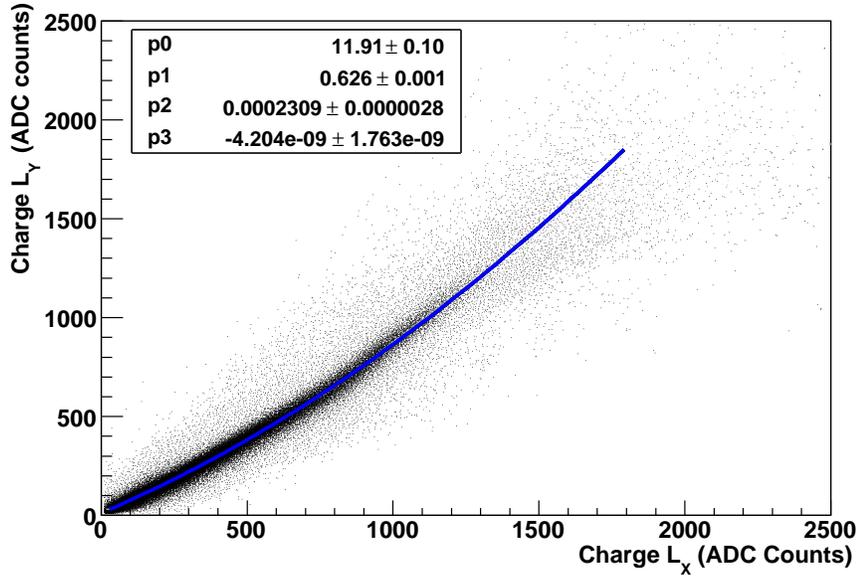}
  \caption{Correlation of induced charges on the cathode planes $x$
    and $y$ in MWPC~L. A third order polynomial function
    is used to parametrise phenomenologically the distribution.}
  \label{fig:chargeLXvsLY}
\end{figure}

As the MWPCs were constructed with symmetric cathode planes, a
correlation exists between the induced charge measured in one cathode
plane to the induced charge of the perpendicular
plane. Fig.~\ref{fig:chargeLXvsLY} shows a large set of measured
charges in MWPC~L deposited by particles of different velocities. A
third order polynomial function was used to parametrise
phenomenologically the distribution. The non-linear response is
attributed to many factors, the asymmetry in the liberated charges
drifting through the two-step MWPC, the systematics in the
cluster-finding, and saturation in the ADCs. In each MWPC the measured
difference of a pair of clusters in $x$- and $y$-direction to the
phenomenological curve is used to determine the quality factors
$Q_x/Q_y|_L$ and $Q_x/Q_y|_M$.  In many cases with these factors alone
a correct pairing of clusters is possible, as shown in
Fig.~\ref{fig:MWPCevent164}.

Valid tracks are bound to angular limits, given by the acceptance of
the spectrometer. Both MWPCs are situated in a field-free space, so
that particle tracks can be extrapolated linearly to the field
boundary at the end of the dipole.

Especially the relation between vertical hit positions of the two
chambers provides a powerful criterion for the track finding. The
target angle acceptance of the spectrometer is large in the horizontal
direction and small in the vertical direction. The vertical track
angles, $\phi$, for particles originating at the target are strongly
correlated to the vertical positions. This $y-\phi$ relation is caused
by the magnet optics leading to diverging particles tracks.

\begin{figure}
 \centering
  \includegraphics[width=0.28\columnwidth]{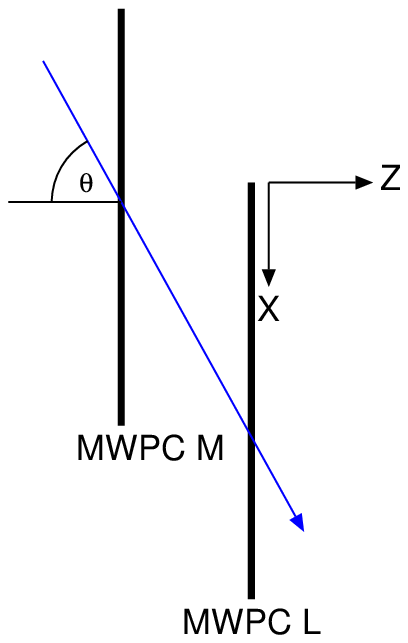}
  \includegraphics[width=0.7\columnwidth]{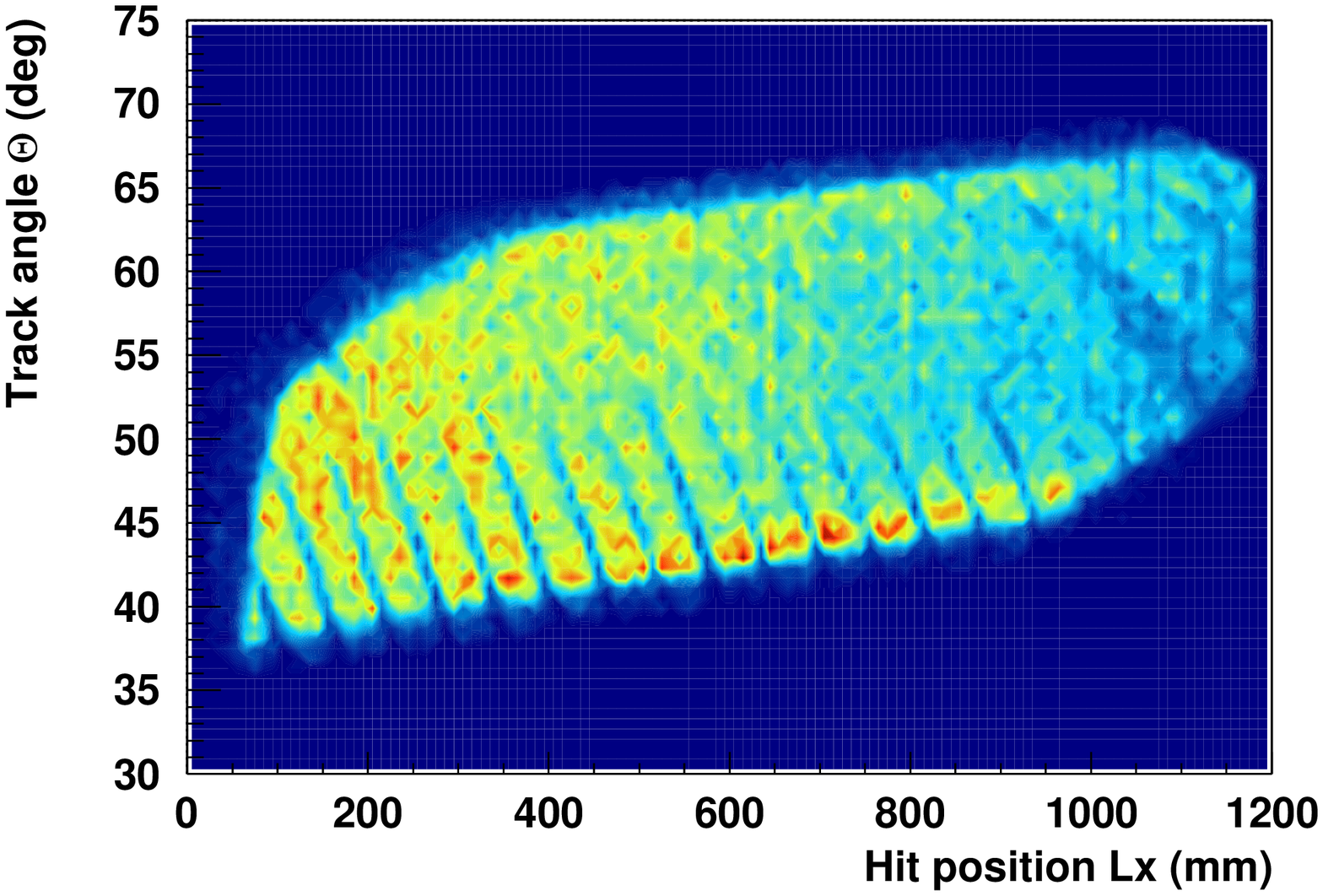}
  \caption{Left: Top view of the MWPCs in the \kaos\ spectrometer with
    the definition of the coordinate system and the horizontal track
    angle. Right: Distribution of events in the $x-\theta$ plane. The
    MWPC acceptance is formed by the bright region. The strip
    structures within the acceptance region are due to the geometry of
    the scintillator walls that leads to inefficiencies.}
  \label{fig:LxvsTheta}
\end{figure}

In the horizontal direction the tracks originating at the target form
a large, but limited acceptance region with horizontal track angles,
$\theta$, ranging from 35 to $\approx 70^{\circ}$, with a dependence
on the horizontal position as shown in Fig.~\ref{fig:LxvsTheta}. The
strip structures within the acceptance region $x-\theta$ are due to
the geometry of the scintillator walls that leads to inefficiencies.

\begin{figure}
  \centering
  \includegraphics[width=0.7\columnwidth]{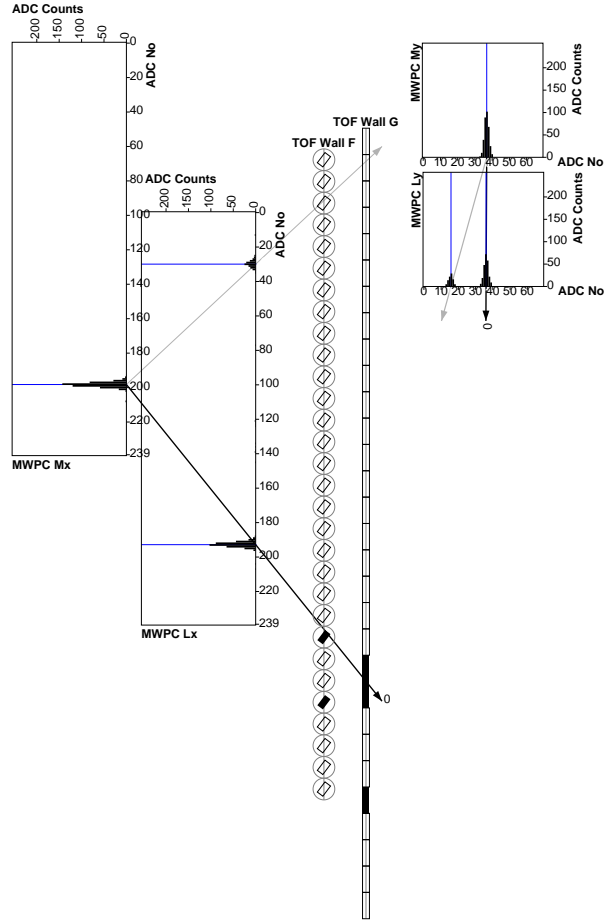}
  \caption{Display of an exceptionally simple event showing in black
    the ADC charges read from the $x$-planes of the MWPCs (left), the
    hits in the scintillator walls $F$ and $G$ (centre) and the ADC
    charges read from the $y$-planes (top right). Particles are
    originating in the top-left corner. One unique set of clusters in
    the MWPCs together with hits in the walls form the track (black
    arrow), the two other track combinations were excluded by the
    quality criteria. The track quality was close to 1, see
    Table~\ref{tab:tracks}.}
  \label{fig:MWPCevent103}
\end{figure}
\begin{table*}
  \caption{Track qualities for the two events displayed in
    Figs.~\ref{fig:MWPCevent103} and \ref{fig:MWPCevent156}. The
    quality factors $Q_x/Q_y$ relate the charges in the two orthogonal
    cathode planes, the quality factors for the positions {\em vs.}
    cartesian angles $x-\theta$ and $y-\phi$ correspond to the
    spectrometer acceptance, the quality factors $\Delta x$ correspond
    to the spatial difference of the projection of the track to the
    hit positions in the scintillator walls $F$ and $G$, and the
    quality factors $\Delta t$ correspond to the spacial difference of
    the projection of the track to the vertical positions as measured
    by the top$-$bottom time difference by the scintillator paddles of
    the walls. It is known that track no.~0 is the proper track, as 
    explained in Section~\ref{sec:efficiencies}.
    \vspace{2mm}}
  \centering
  \begin{tabular}{cc c cc cc cc cc}
    \hline
    event & track & total & $Q_x/Q_y|_M$ & $Q_x/Q_y|_L$ &
    $x-\theta$ & $y-\phi$ & $\Delta x|_F$ & $\Delta x|_G$ &
    $\Delta t|_F$ & $\Delta t|_G$\\
    \hline
    \hline
    Fig.~\ref{fig:MWPCevent103}
    & 0 & 0.96 & 1    & 1    & 1 & 1 & 1    & 0.96 & 0.58 & 1\\
    \hline
    Fig.~\ref{fig:MWPCevent156}
    & 0 & 1    & 1    & 1    & 1 & 1 & 1    & 1    & 0.81 & 1\\
    & 1 & 0.73 & 0.86 & 0.86 & 1 & 1 & 1    & 1    & 0.81 & 1\\
    & 2 & 0.40 & 0.55 & 1    & 1 & 1 & 0.73 & 1    & 0.91 & 1\\   
    \hline
  \end{tabular}
  \label{tab:tracks}
\end{table*}
\begin{figure}
  \centering
  \includegraphics[width=0.7\columnwidth]{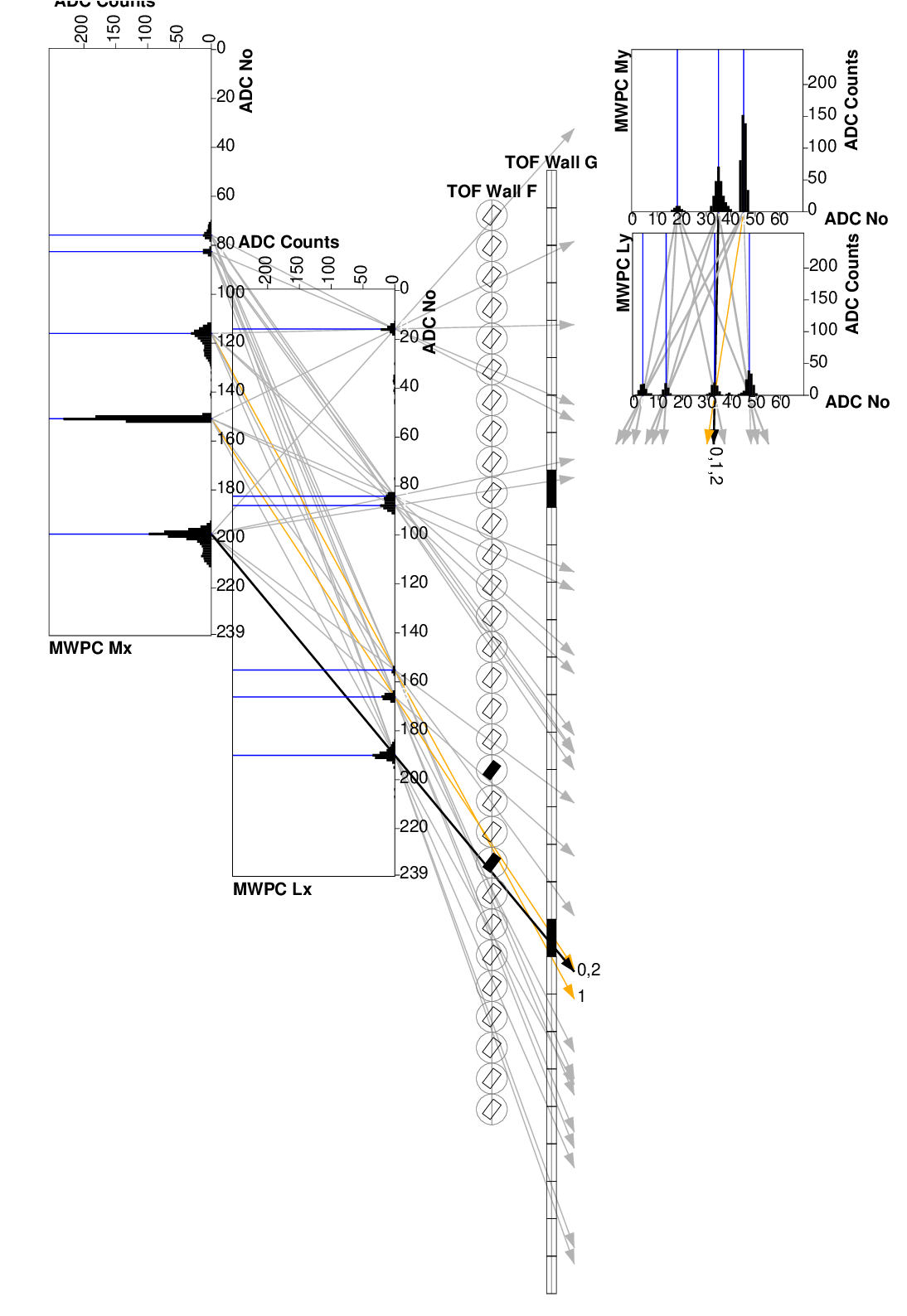}
  \caption{Same as Fig.~\ref{fig:MWPCevent103} but for a more common
    event topology with multiple clusters or hits in each detection
    plane. From the charge distribution in the $x$-plane of MWPC~M the
    coincident passing of one heavily ionizing and at least one
    minimum ionizing particle through this chamber can be deduced. The
    track-finding algorithm returned three possible tracks (accentuated
    arrows, listed in Table~\ref{tab:tracks}), all other
    cluster and hit combinations were excluded by at least one of the
    quality criteria.}
  \label{fig:MWPCevent156}
\end{figure}

In the \kaos\ spectrometer the MWPCs are followed by two scintillator
walls, labelled F and G, that are segmented in $x$-direction.  A last
group of quality factors is determined by the extrapolation of the
track to both scintillator walls. The spatial difference to the
closest observed hit, $\Delta x|_F$ and $\Delta x|_G$, are used as
well as the the spacial difference to the vertical positions, $\Delta
t|_F$ and $\Delta t|_G$, as measured by the top$-$bottom time
differences by the individual scintillator paddles.  The measured time
difference in a scintillator bars becomes less accurate when the
particle deposits only a small amount of energy, {\em e.g.}\ for
grazing incidence. To improve the quality assessment the maximum value
from both walls is used as quality factor.

Figs.~\ref{fig:MWPCevent103} and \ref{fig:MWPCevent156} show the event
displays of two events, the first being exceptionally simple with a
low occupancy of ADC charges and hits in the scintillator walls. It
can be seen in this event that a strong charge correlation between the
two peaks in $x$- and the two peaks in $y$-direction exits. One unique
set of clusters in the MWPCs together with hits in the walls form the
track, the two other track combinations were excluded by the quality
criteria and assigned to background particles. The track quality was
close to 1, see Table~\ref{tab:tracks}. A more common event topology
with multiple clusters or hits in each detection plane is shown in
Fig.~\ref{fig:MWPCevent156}. From the charge distribution in the
$x$-plane of MWPC~M the coincident passing of one heavily ionizing and
at least one minimum ionizing particle through this chamber can be
deduced. The track-finding algorithm returns three possible tracks see
Table~\ref{tab:tracks}), all other cluster and hit combinations were
excluded by one of the quality criteria. As both events were triggered
by the efficiency counters, explained in
Section~\ref{sec:efficiencies}, it is known that with track no.~0 a
proper track assignment was achieved.
				
\section{Tracking Efficiencies}
\label{sec:efficiencies}

The implemented read-out principle allows for high track
multiplicities, however, the track reconstruction can then become a
major source of detection inefficiency, quantified by the tracking
efficiencies as follows: (i) intrinsic efficiency: the percentage of
events in which any charge was detected in the chamber following a
charged particle, (ii) any track efficiency: the percentage of events
in which a track was reconstructed from the charges, (iii) track
reconstruction efficiency: the percentage of events in which the
proper track was reconstructed.  All efficiencies are to some extend
depending on the cluster analysis and the reconstruction method. The
track reconstruction efficiency is heavily depending on the particle's
momentum and the beam current, or more general on the luminosity. The
latter dependency is a consequence of the fast increment of the flux
of background particles with the current.

There is no way of determining the tracking efficiencies from the
electro-production measurements. Instead, data was taken during
dedicated efficiency runs.  For these runs two small scintillating
detectors of type Bicron BC-408 and dimensions $L \times W \sim 30
\times 20$\,mm$^2$ with 5\,mm thickness were installed in front of
each MWPC to determine intrinsic and tracking efficiencies of the
chambers. The active parts were connected to optical light guides and
read out by PMTs of type Hamamatsu R1828. When attached to the MWPC
frames the active areas of the counters were vertically centred and
horizontally at the same position relative to the
corresponding MWPC. The detectors could easily be moved out of the
spectrometer's acceptance when not in use.

\begin{figure}
  \centering
  \includegraphics[width=0.8\columnwidth]{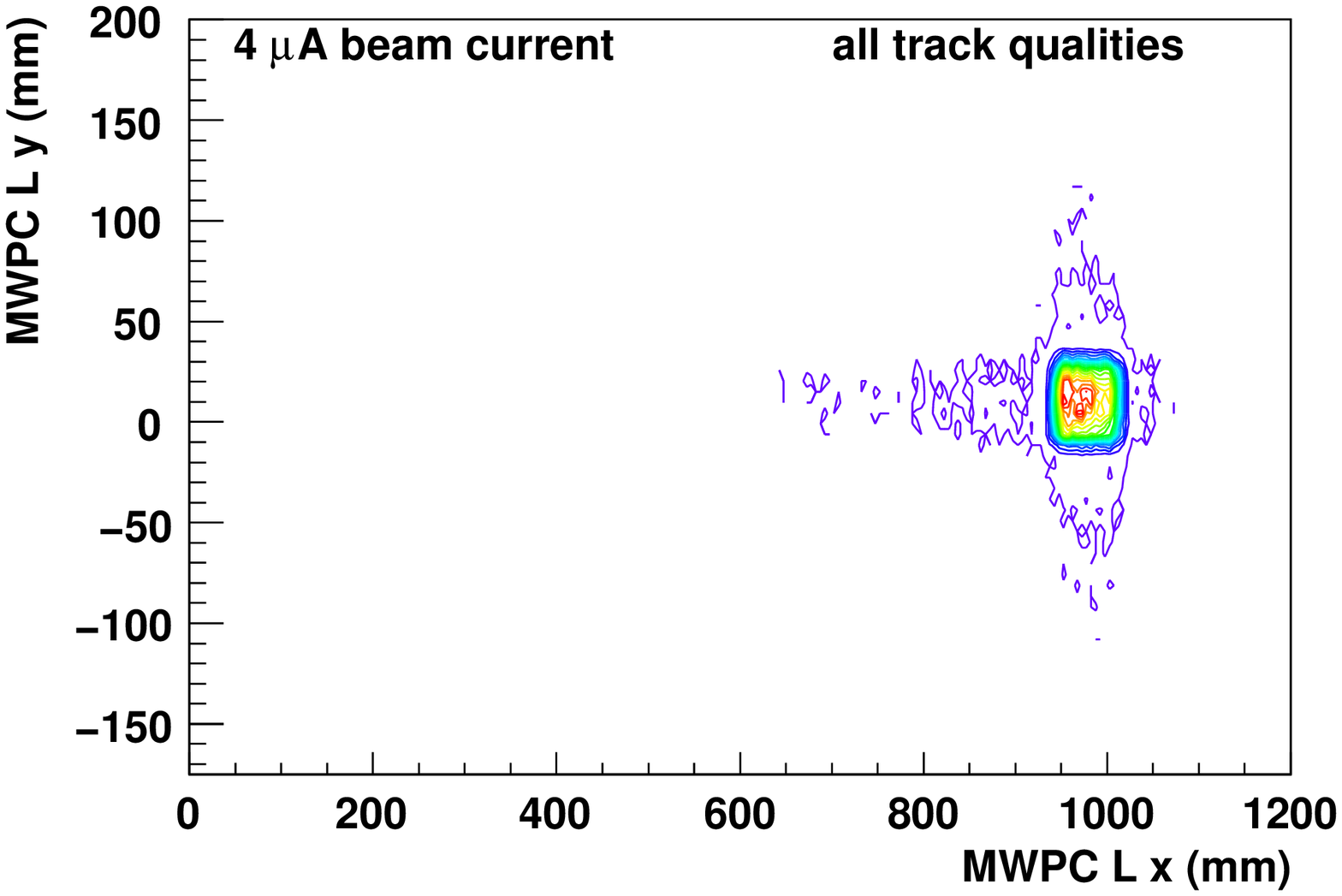}
  \includegraphics[width=0.8\columnwidth]{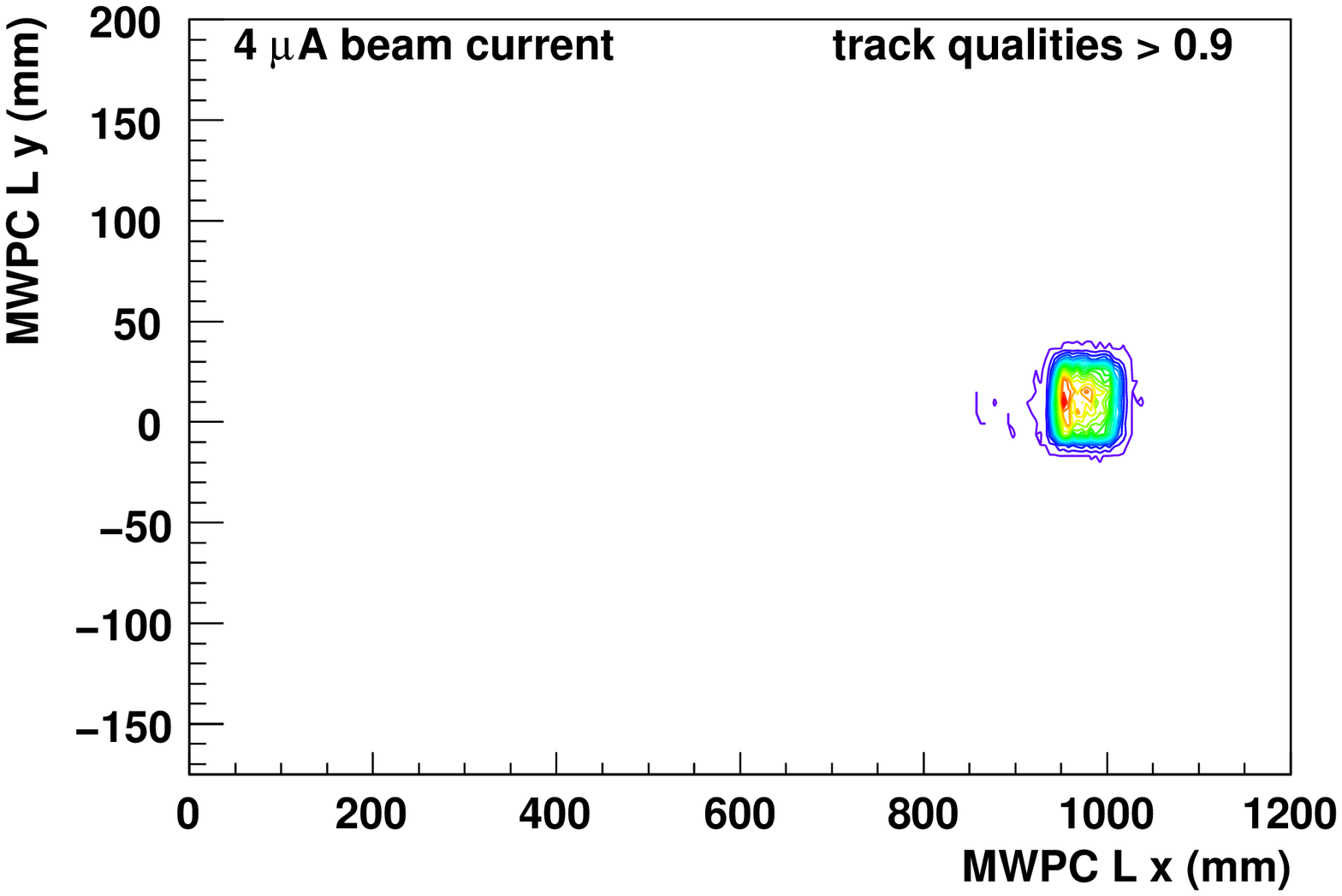}
  \caption{Reconstructed track positions in MWPC~L for events
    triggered by the efficiency counters at a beam current of
    4\,$\mu$A for all track qualities (top) and for track qualities
    larger 0.9 (bottom). The densely populated square (approx.\ 93
    $\times$ 58 mm$^2$) corresponds to the projection of the position
    of the efficiency counters to the wire chamber coordinate system
    which defines the acceptance region of this chamber. Tracks
    found to be outside of this acceptance region were reconstructed 
    wrongly. These tracks tend to have a low track quality.}
  \label{fig:Lxy}
 \end{figure}

The external counters could be used to select trajectories passing
through both chambers in a small acceptance region. The histogram in
Fig.~\ref{fig:Lxy} shows a typical distribution of events in the wire
chamber coordinate system obtained under those trigger conditions for
a beam current of 4\,$\mu$A, corresponding to luminosities of 5
$\times$ 10$^{36}$\,cm$^2$s$^{-1}$. The bands in $x$- and
$y$-direction correspond to wrongly reconstructed coordinates in one
of the four MWPC planes, the strongly populated square corresponds to
the projection of the position of the efficiency counters. When
requiring a high track quality almost only properly reconstructed
tracks remain in the event sample.

\begin{figure}
  \centering
  \includegraphics[width=\columnwidth]{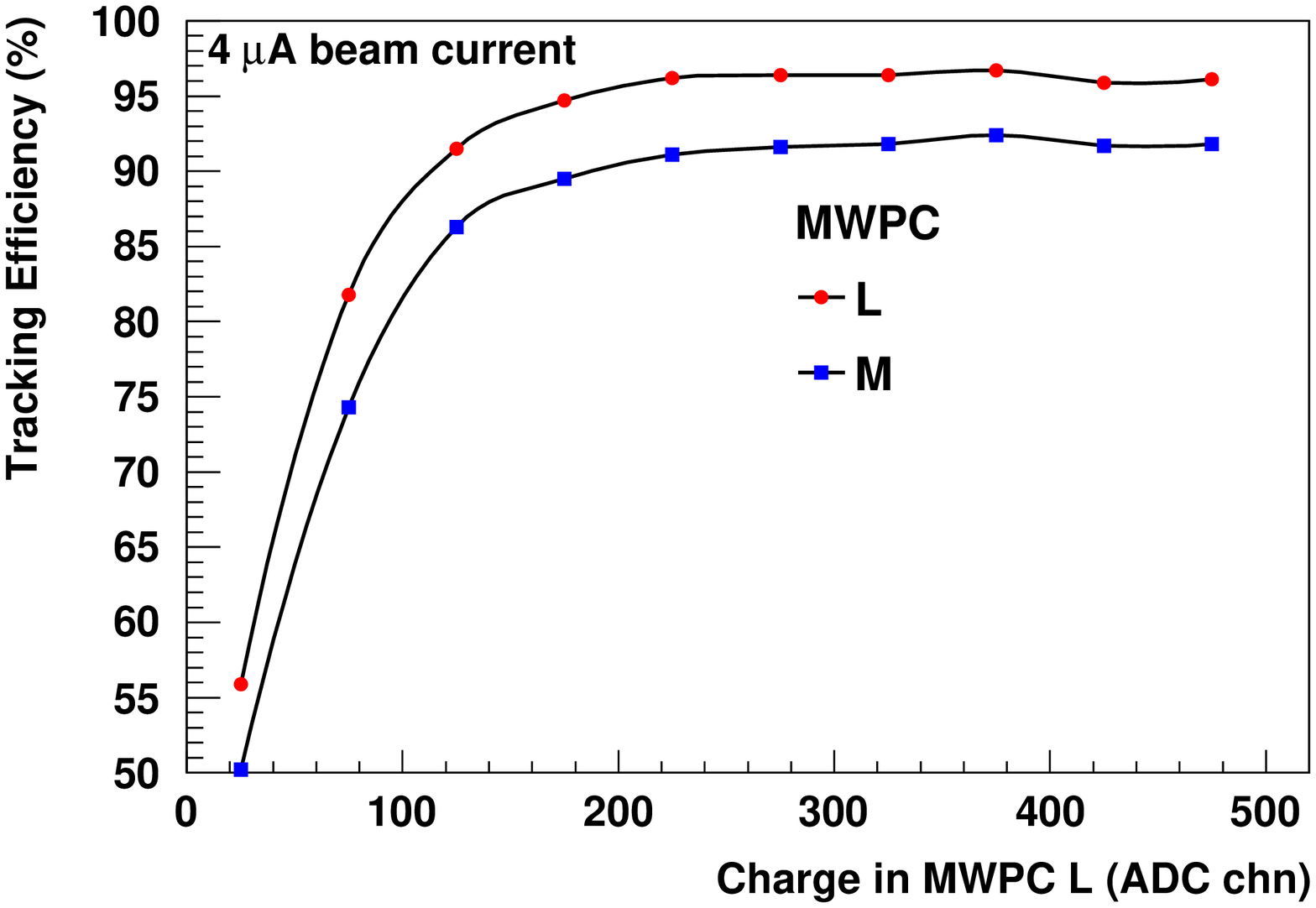}
  \caption{Charge-separated track reconstruction efficiencies for a
    beam current of 4\,$\mu$A. Efficiencies for both chambers are
    shown as a function of induced charge on the cathode of MWPC
    L. The average charge collected in chamber L for pions is $\sim$
    65 ADC counts, and for protons $\sim$ 220 ADC counts.}
  \label{fig:ADC-separated-Eff}
\end{figure}

The spectrometer was situated under forward angles at 31$^\circ$ with
a solid angle acceptance of $\Omega \approx$ 10\,msr in an in-plane
angular range of $\vartheta = 21 - 43^\circ$.  The results are
separated into samples taken with 1, 2, 3, and 4\,$\mu$A beam current
and within the samples into events where the traversing particle was
identified by its specific energy-loss as proton and pion by the
scintillator walls.  With the spectrometer set to a momentum bite
around 600\,MeV$\!/c$ protons, kaons, and pions loose significantly
different energies in the counter gas. The intrinsic efficiencies were
well above 99\,\% independent on the beam current. The efficiency to
form a track was somewhat above 98\,\%. For pions track reconstruction
efficiencies ranged between 86 and 68\,\% corresponding to beam
currents of 1--4\,$\mu$A, for protons the efficiencies were between 98
and 90\,\%, details are tabulated in
Table~\ref{tab:efficiencies}. Several runs were taken during a period
of 2 weeks showing small variations of no more than
2\,\%. Charge-separated tracking efficiencies for a beam current of
4\,$\mu$A are shown in Fig.~\ref{fig:ADC-separated-Eff}. From the
average charge collected for kaons track reconstruction efficiencies
of 75--90\,\% were evaluated.

\begin{table*}
  \centering
  \caption{Intrinsic particle detection and tracking efficiencies as
    determined with the efficiency counter set-up. A track is defined
    as being properly reconstructed if it is found to be in the
    acceptance region of the respective chamber. The gain was higher
    in chamber L than in chamber M, so that tracking efficiencies
    differ by a few percent, this difference is increasing with beam
    current. The chamber track reconstruction efficiencies in columns 5
    and 6 include all track qualities from 0 to 1 and all particle
    species. These efficiencies are separated into pion and proton tracks.
    Larger inefficiencies arise for higher beam currents and for
    pion tracks due to the resulting ambiguities and
    distortions in the charge distributions.
    The bottom 3 lines show the loss of efficiency when not all quality
    factors are used in the track reconstruction.\vspace{2mm}}
  \begin{tabular}{cc | ccccccccc l}
    \hline
    $\cal L$/10$^{36}$& $I$ & \multicolumn{2}{c}{intrinsic $\epsilon$ (\%)} & 
    \multicolumn{3}{c}{tracking $\epsilon$ (\%)} &
    \multicolumn{2}{c}{proton $\epsilon$ (\%)} &
    \multicolumn{2}{c}{pion $\epsilon$ (\%)}\\
    (cm$^2$s$^{-1}$) & ($\mu$A) & L & M & any track & L & M & L & M & L & M & 
    used quality factors\\
    \hline
    1.2 & 1 & 99.3 & 99.6 & 97.8 & 97.1 & 95.9 & 98.3 & 97.1 & 86.4 & 83.9 & all\\
    2.5 & 2 & 99.5 & 99.7 & 97.6 & 96.0 & 93.4 & 97.4 & 95.1 & 83.5 & 78.4 & ''\\
    3.7 & 3 & 99.6 & 99.8 & 97.3 & 94.5 & 90.9 & 96.0 & 92.7 & 81.7 & 74.9 & ''\\
    5 & 4 & 99.6 & 99.8 & 97.3 & 93.1 & 88.2 & 95.0 & 90.6 & 76.9 & 67.9 & ''\\
    \hline
    5 & 4 &  ''  &  ''  & 98.9 & 75.8 & 62.7 & 78.2 & 65.5 & 54.3 & 39.0
    & no $\Delta x|_{F,G}$, no $\Delta t|_{F,G}$\\
    5 & 4 &  ''  &  ''  & 97.8 & 90.6 & 80.8 & 92.5 & 83.8 & 74.0 & 55.8
    & no $x-\theta$, no $y-\phi$\\
    5 & 4 &  ''  &  ''  & 97.9 & 91.4 & 83.7 & 93.1 & 85.8 & 77.1 & 66.5
    & no $Q_x/Q_y|_{M,L}$\\
    \hline
  \end{tabular}
  \label{tab:efficiencies}
\end{table*}

\begin{figure}
  \centering
  \includegraphics[width=\columnwidth]{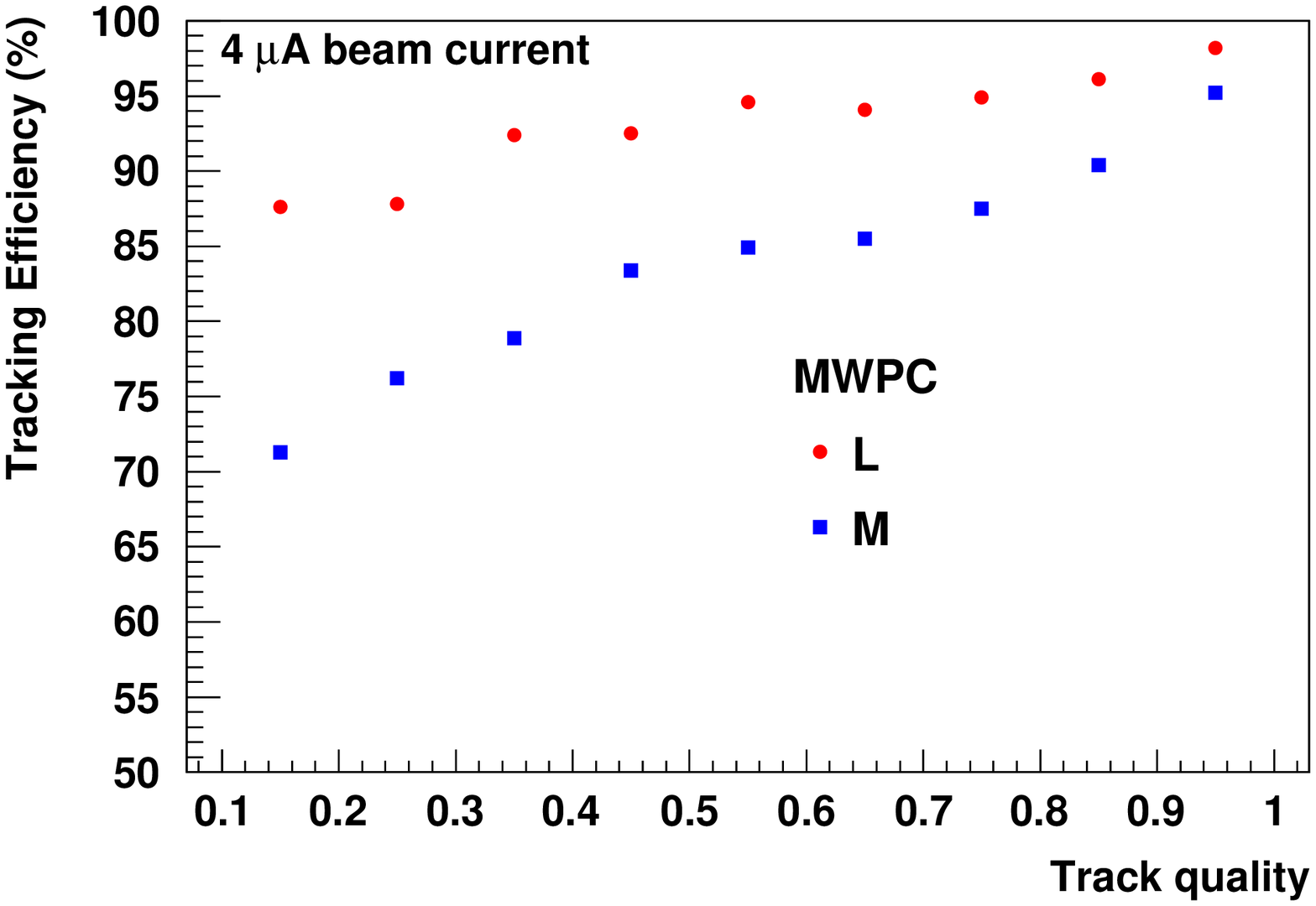}
  \caption{Quality-separated track reconstruction efficiencies for a
    beam current of 4\,$\mu$A. Efficiencies for both chambers are
    shown as a function of a quality factor that includes, among
    others, the correlation that exists between the induced charge
    measured in one cathode plane to the induced charge of the
    perpendicular plane and the relation between vertical hit
    positions in the two chambers. The continuous increase of the
    efficiency with the quality provides confidence that the quality
    factor is indeed a measure of the goodness-of-track.}
  \label{fig:q-separated-Eff}
\end{figure}

Tracks reconstructed to pass through the acceptance region on average
have a higher quality factor, whereas tracks reconstructed to miss
this region on average have a smaller quality factor. The continuous
increase of the efficiency with the quality as shown in
Fig.~\ref{fig:q-separated-Eff} for 4\,$\mu$A beam current, provides
confidence that the quality factor is indeed a measure of the
goodness-of-track.

%

\section{Discussion}
\label{sec:discussion}

When using the cathode-charge sampling method for the bi-dimensional
read-out of MPWCs a large variation of induced charge distributions
can be observed, depending {\em e.g.}\ on track angles, charged
particle velocities, integration gate width and timing, and charged
particle occupancies. For MWPCs used as tracking detectors in a
magnetic spectrometer these factors vary, as incident particle species
and momenta as well as background particle fluxes depend on the
spectrometer and accelerator settings for a particular experiment.

To resolve the ambiguities appearing with multiple, non-perpendicular
incident particles with different energy-losses a cluster-finding and
track reconstruction algorithm was developed that assigns each
possible track a quality factor. This method was in-beam tested with
two large-sized MWPCs in the \kaos\ spectrometer and proved to be
effective not only in quantifying the tracking efficiencies, but more
importantly it allowed to systematically determine track
reconstruction efficiencies which can be applied to the operation of
the MWPCs in electro-production experiments.
 
In general, tracks of heavily ionising particles in a low background
environment have the largest probability of being properly
reconstructed. Tracks of minimum ionizing particles are harder to
reconstruct properly as background signals of dominantly small
amplitudes are present in the chambers at any time. It is concluded
that the track reconstruction in the MWPCs represents a primary source
of inefficiency for high electron beam currents on typical targets of
several hundred mg$/$cm$^2$ thickness.  The use of information from
the external detectors improves the efficiencies by 20--25\,\% in the
case of the highest beam current that was probed. However, at these
currents also the external counters register many multiply hits. The
check on the spectrometer acceptance and the correlations of the
coordinates improves the efficiencies by 10--15\,\%, and the induced
cathode charge correlation has an impact of $\sim$ 5\,\% for the
higher amplitudes.

By using the introduced method and the obtained results the track
reconstruction efficiencies for particles not seen in the beam-tests,
{\em i.e.}\ kaons, could have been determined. These findings were
applied to data taken with the \kaos\ spectrometer in kaon
electro-production, improving the systematic uncertainties in the
cross-section extraction significantly.

\section*{Acknowledgements}
%
Work supported in part by the Federal State of Rhine\-land-Palatinate
and by the Deutsche Forschungsgemeinschaft with the Collaborative
Research Center 443.



\end{document}